\documentclass[fleqn,usenatbib]{mnras}
\usepackage{newtxtext,newtxmath}
\usepackage{graphicx,amsmath,booktabs,bm,url,placeins}
\title[Frozen composition and rotational crust failure]{Frozen-composition effects on rotational failure of neutron-star crusts}
\author[E. Giliberti]{E. Giliberti}
\date{6 September 2026}
\pubyear{2026}
\begin{document}\label{firstpage}\pagerange{\pageref{firstpage}--\pageref{lastpage}}\maketitle

\begin{abstract}
A changing rotation rate loads a neutron-star crust because the preferred fluid equilibrium figure evolves while the solid resists deformation. The resulting strain depends not only on the shear modulus but also on the compressional response of compositionally stratified matter. We quantify the difference between chemical equilibrium, $K_{\rm eq}=n_{\rm B}dP_{\rm eq}/dn_{\rm B}$, and the frozen-composition limit, $K_f=n_{\rm B}(\partial P/\partial n_{\rm B})_{\{Y_i\}}$, using realistic BSk24, BSk21 and SLy4 inner-crust microphysics in an axisymmetric Newtonian hydro-elastic model. In the matched $0.01\le n_{\rm B}\le0.06~{\rm fm^{-3}}$ domain, frozen composition lowers the rotational failure spin by $16.7$--$21.7$ per cent for a $1.4\,M_\odot$ star. The effect is not a uniform rescaling of the strain: for BSk24 the first-failure layer moves from $n_{\rm B}\simeq0.0284$ to $0.0361~{\rm fm^{-3}}$, and the direction of the shift is EoS dependent. Density-resolved perturbations show that the global change in failure threshold is controlled by the radial overlap between the microphysical frozen stiffening and the mechanical susceptibility of the crust; for BSk21/24 about $82$ per cent of the linear sensitivity arises from $0.01<n_{\rm B}<0.03~{\rm fm^{-3}}$. This provides a microphysical realisation of the non-barotropic rotational response explored previously with phenomenological adiabatic indices and shows that composition freezing changes both how much strain is supported and where the crust becomes vulnerable.
\end{abstract}

\section{Introduction}
The solid crust of a neutron star stores elastic stress generated by changes in rotation, magnetic forces, accretion, superfluid dynamics and external tides. This makes crustal elasticity relevant to starquakes and glitches \citep{BaymPines1971,FrancoLinkEpstein2000}, magnetar activity \citep{Lander2015}, tidal crust failure \citep{Tsang2012}, and the maximum non-axisymmetric deformation that can source continuous gravitational waves \citep{Ushomirsky2000,HaskellJonesAndersson2006,JohnsonMcDanielOwen2013,GittinsAnderssonJones2021}. Molecular-dynamics and lattice calculations indicate that Coulomb matter can be unusually strong, although the effective breaking threshold depends on crystal structure, temperature, strain rate and deformation geometry \citep{HorowitzKadau2009,ChugunovHorowitz2010,Baiko2018,CaplanHorowitz2017}.

Rotation is a particularly useful loading problem because it separates the global forcing from the constitutive response. As the spin changes, the preferred hydrostatic figure changes. A fluid follows that sequence without shear stress; a solid crust cannot, and therefore develops an $\ell=0,2$ displacement and a spatially non-uniform strain field. Classical starquake models already exploited this mismatch between the evolving equilibrium figure and an elastic crust \citep{BaymPines1971,FrancoLinkEpstein2000}. More recent work has developed increasingly realistic elastic boundary-value formulations, including the treatment of gravitational-potential perturbations and crust--fluid interfaces \citep{HaskellJonesAndersson2006,Giliberti2020,GittinsAnderssonJones2021}.

A second piece of physics is equally important: the background star is in chemical equilibrium, but a perturbed fluid element need not remain so. If weak or nuclear reactions are slow compared with the loading, composition is approximately frozen along the displacement. The perturbed pressure then depends on density at fixed composition rather than along the equilibrium sequence, producing a non-barotropic response and stable compositional stratification \citep{ReiseneggerGoldreich1992,ChamelHaensel2008}. In a previous paper, \citet{Giliberti2020} incorporated this physics phenomenologically by allowing the adiabatic index of the perturbation to differ from that of the equilibrium polytrope. That study showed that even a modest change of the perturbative adiabatic index can strongly modify rotation-induced displacements and strains.

The present work asks the next, more microphysical question. Instead of prescribing a phenomenological frozen adiabatic index, we reconstruct the local frozen bulk modulus from realistic inner-crust composition for BSk24, BSk21 and SLy4. We then determine how this EoS-dependent constitutive change modifies the \emph{failure threshold} and the \emph{location of first failure}. The central issue is therefore not the absolute Newtonian breaking frequency, but how a realistic composition-dependent $K_f-K_{\rm eq}$ is filtered through the non-uniform mechanical susceptibility of the crust.

A recent finite-element calculation by \citet{MoralesHorowitz2025} provides a useful independent benchmark for rotational crust failure, finding first failure near $0.58\,\Omega_K$ in a Newtonian $n=1$ polytrope with a simplified shear law. Because the stellar background, shear prescription and elastic treatment differ, we do not use that absolute value as a calibration target. Its role here is contextual: our focus is the differential constitutive effect of chemical freezing within a fixed stellar and mechanical model.

This distinction leads to three questions. How large is the failure-frequency shift when $K_f$ is computed from realistic crust composition? Does frozen composition merely rescale the strain, or does it move the vulnerable layer? And why do different EoSs respond differently? Numerical and robustness details that do not bear directly on these physical questions are collected in the appendices.

\section{Physical framework}\label{sec:model}
\subsection{Equilibrium and frozen compression}
For cold catalysed matter the stellar background follows an equilibrium barotrope $P_{\rm eq}(n_{\rm B})$, with
\begin{equation}
K_{\rm eq}=n_{\rm B}\frac{dP_{\rm eq}}{dn_{\rm B}}.
\end{equation}
A displaced element whose composition cannot readjust instead responds at approximately fixed particle fractions,
\begin{equation}
K_f=n_{\rm B}\left(\frac{\partial P}{\partial n_{\rm B}}\right)_{\{Y_i\}}.
\end{equation}
The difference $K_f-K_{\rm eq}$ is the local microphysical source of the effect studied here. The background remains fixed; only the constitutive relation of the perturbation changes. This is essential for interpreting the frozen/equilibrium ratio as a differential response.

The frozen limit is the natural high-frequency or slow-reaction limit of a compositionally stratified medium \citep{ReiseneggerGoldreich1992}. In the crust, the same principle implies that the perturbative bulk response need not follow the equilibrium sequence \citep{ChamelHaensel2008}. Here this distinction is instead supplied by an EoS-dependent radial profile of the frozen compressional response.

\subsection{Rotational loading and elastic failure}
We use a cold spherical Newtonian background and perturb it by a change in centrifugal potential. The displacement contains $\ell=0$ and $\ell=2$ components, while the isotropic stress is closed by $\Delta P=-K\nabla\cdot\bm\xi$ and the deviatoric stress by the shear modulus $\mu$. The fluid gravitational response $W_\ell=\delta\Phi_\ell+\chi_\ell$ is computed once and held fixed in every matched equilibrium/frozen pair. The full radial system, interface conditions and strain reconstruction are given in Appendix~\ref{app:odes}.

Because the linear solution scales with $\Delta\Omega^2$, the strain invariant scales as $\Omega^2$. For a reference solution at $\Omega_*$,
\begin{equation}
\Omega_b=\Omega_*\left[\frac{\varepsilon_{\rm crit}}{\varepsilon_{\max}(\Omega_*)}\right]^{1/2}.
\end{equation}
We use a von-Mises threshold $\varepsilon_{\rm crit}=0.1$, motivated by Coulomb-crystal simulations \citep{HorowitzKadau2009}. Alternative failure and shear prescriptions are supporting systematics and are moved to Appendix~\ref{app:robust}. The primary quantity is $\Omega_b^f/\Omega_b^{\rm eq}$, not the absolute Newtonian value of $\Omega_b/\Omega_K$.

\subsection{Realistic crust microphysics}\label{sec:micro}
The production calculations use the unified BSk24 and BSk21 descriptions \citep{Pearson2012,Pearson2018} and the unified SLy4 model \citep{DouchinHaensel2001,HaenselPotekhin2004}. These models provide different equilibrium composition gradients and therefore different radial structures of $K_f/K_{\rm eq}$. The elastic baseline is a composition-dependent Coulomb-lattice shear modulus,
\begin{equation}
\mu_V=0.1194\,n_i\frac{(Ze)^2}{a},\qquad a=\left(\frac{3}{4\pi n_i}\right)^{1/3},
\end{equation}
consistent with the standard picture of a strongly coupled Coulomb crystal \citep{ChamelHaensel2008,CaplanHorowitz2017}. Details of the pressure-renormalised frozen reconstruction, thermodynamic floor and polycrystalline audit are given in Appendix~\ref{app:numerics}.

\begin{figure}
\centering
\includegraphics[width=0.98\columnwidth]{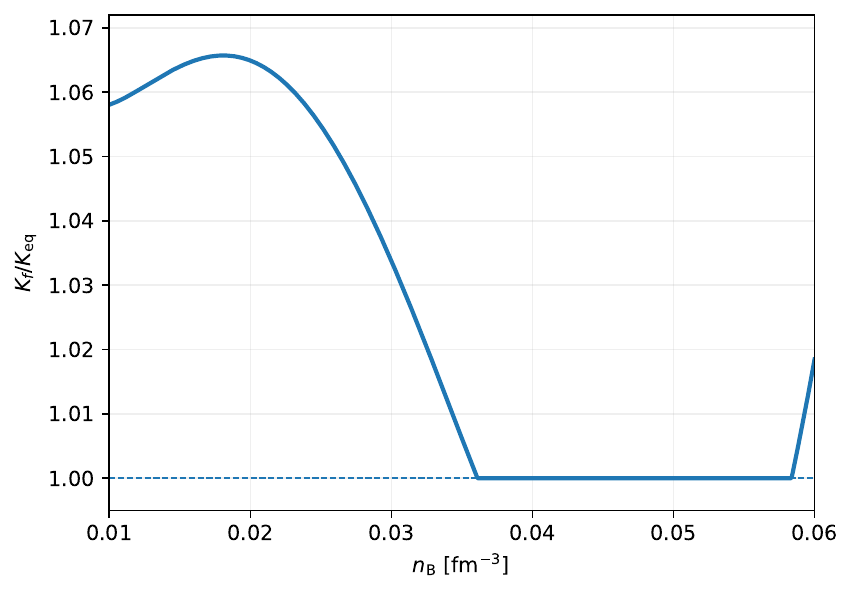}
\caption{Frozen-to-equilibrium bulk-modulus profile for BSk24 in the controlled inner-crust domain, reconstructed from the archived vector data of the production figure. The curve is shown as a spatial constitutive profile, not as a scalar measure of the total failure shift.}
\label{fig:kf}
\end{figure}

The controlled comparison is restricted to $0.01\le n_{\rm B}\le0.06~{\rm fm^{-3}}$. Figure~\ref{fig:kf} shows the archived BSk24 frozen-to-equilibrium profile recovered from the vector data of the production figure. We deliberately show only the profile for which a pointwise archival reconstruction is available, rather than mixing it with schematic reconstructions for the other EoSs. The relevant information is not simply the peak value of $K_f/K_{\rm eq}$, but where the stiffening is located relative to the elastic response of the star.

\section{Results: how composition freezing changes failure}\label{sec:results}
\subsection{Frozen composition amplifies and redistributes strain}
The first effect of composition freezing is quantitative, but the more revealing effect is spatial. For BSk24 the certified Voigt calculation gives $\Omega_b^{\rm eq}=0.4997\Omega_K$ and $\Omega_b^f=0.3912\Omega_K$, hence $\Omega_b^f/\Omega_b^{\rm eq}=0.783$ and a $21.7$ per cent reduction. More importantly, in both closures the maximum remains on the equator, while its radial location moves from $n_{\rm B}\simeq0.0284$ to $0.0361~{\rm fm^{-3}}$. Frozen composition therefore changes both the amplitude and the spatial organisation of the strain. The global boundary-value problem reorganises where the crust carries the largest deviatoric load; the effect is therefore more than a multiplicative correction to an effective adiabatic index.
\begin{figure}\centering\includegraphics[width=0.98\columnwidth]{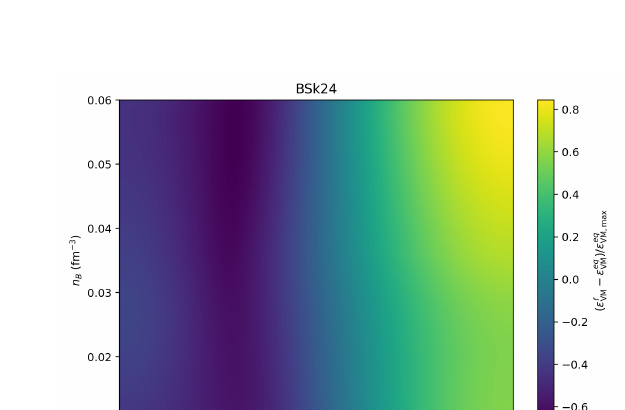}\caption{Change in the BSk24 rotational von-Mises strain field. Frozen composition redistributes the strain rather than acting as a uniform multiplicative correction.}\label{fig:strain}\end{figure}

\begin{table}\centering\caption{Matched-bulk $1.4\,M_\odot$ results.}\label{tab:eos}\begin{tabular}{lcccc}\toprule EoS&$\Omega_b^{\rm eq}/\Omega_K$&$\Omega_b^f/\Omega_K$&ratio&reduction\\\midrule BSk24&0.4997&0.3912&0.783&21.7\%\\ BSk21&0.4972&0.3895&0.783&21.7\%\\ SLy4&0.4699&0.3916&0.833&16.7\%\\\bottomrule\end{tabular}\end{table}

The location of first failure is not universal: frozen composition moves the maximum inward for BSk21/24 but slightly outward for SLy4. The sign of the frequency correction is common to all three EoSs, whereas the spatial migration of the maximum retains detailed information about the underlying composition profile.

\subsection{Dependence on stellar mass}\label{sec:mass}
\begin{figure}
\centering
\includegraphics[width=0.98\columnwidth]{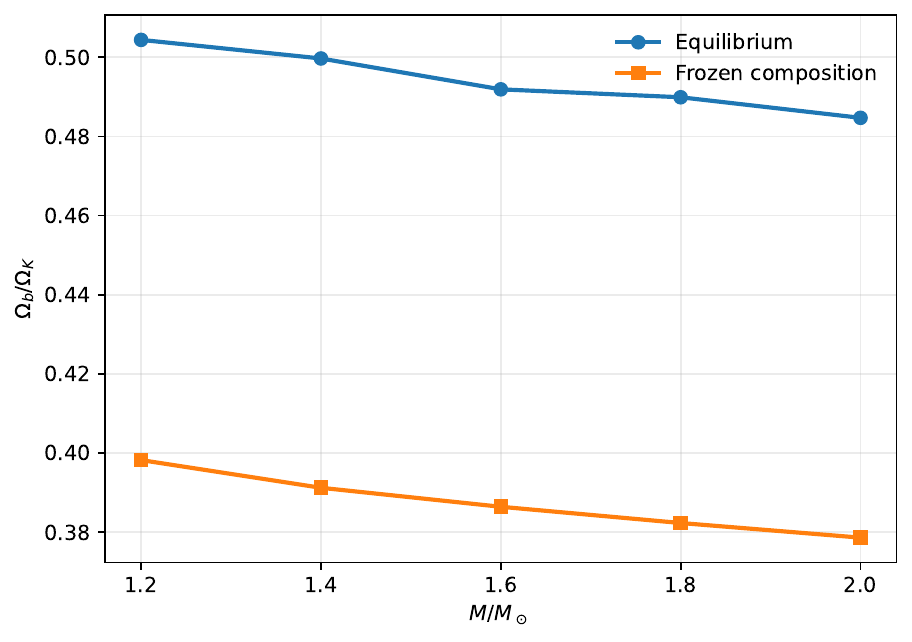}
\caption{BSk24 rotational failure frequency as a function of stellar mass. Equilibrium and frozen-composition sequences remain nearly parallel: the absolute Newtonian threshold decreases with mass, while the frozen-composition reduction remains close to $21$--$22$ per cent.}
\label{fig:massscan}
\end{figure}

The constitutive correction is not restricted to the canonical $1.4\,M_\odot$ model. Figure~\ref{fig:massscan} shows the certified BSk24 sequence from $1.2$ to $2.0\,M_\odot$. The absolute Newtonian failure frequency decreases with mass in both closures, but the equilibrium and frozen sequences remain nearly parallel. Consequently, the fractional reduction stays in the narrow interval $21.05$--$21.97$ per cent across the full scan. This is a useful physical result: within the present Newtonian family, changing the stellar mass alters the absolute rotational scale more strongly than it alters the \emph{relative} penalty associated with composition freezing. We therefore interpret the near-constant separation as evidence that the frozen correction is a constitutive property of the stratified crust rather than a peculiarity of the reference mass. The absolute mass trend itself should not be extrapolated beyond the present Newtonian backgrounds.

\subsection{Why the EoSs differ: radial overlap}
In density band $i$ we scale only the physical frozen-minus-equilibrium correction,
\begin{equation}\Gamma(r;\eta_i)=\Gamma_{\rm eq}(r)+\eta_iW_i(r)[\Gamma_f(r)-\Gamma_{\rm eq}(r)],\end{equation}
and define ${\cal S}_i=-\partial\ln\Omega_b/\partial\eta_i|_0$. For BSk21 and BSk24 about $82$ per cent of the linear frozen sensitivity comes from $0.01<n_{\rm B}<0.03~{\rm fm^{-3}}$. Summing the band-by-band linear responses reconstructs the corresponding whole-profile linear response to about one per cent for the two BSk models, providing a direct check that the localisation is not an artefact of the chosen binning. A second experiment imposes the same local one-per-cent increase in $\Gamma$ in every band; this isolates a mechanical susceptibility that grows inward. The total shift is therefore controlled by the radial overlap of an EoS-specific microphysical source with a non-uniform mechanical response. A layer contributes strongly only if the composition is sufficiently stiffened there and the elastic boundary-value problem is sensitive to compression in that layer. A large local $K_f/K_{\rm eq}$ is not sufficient by itself.
\begin{figure}\centering\includegraphics[width=0.98\columnwidth]{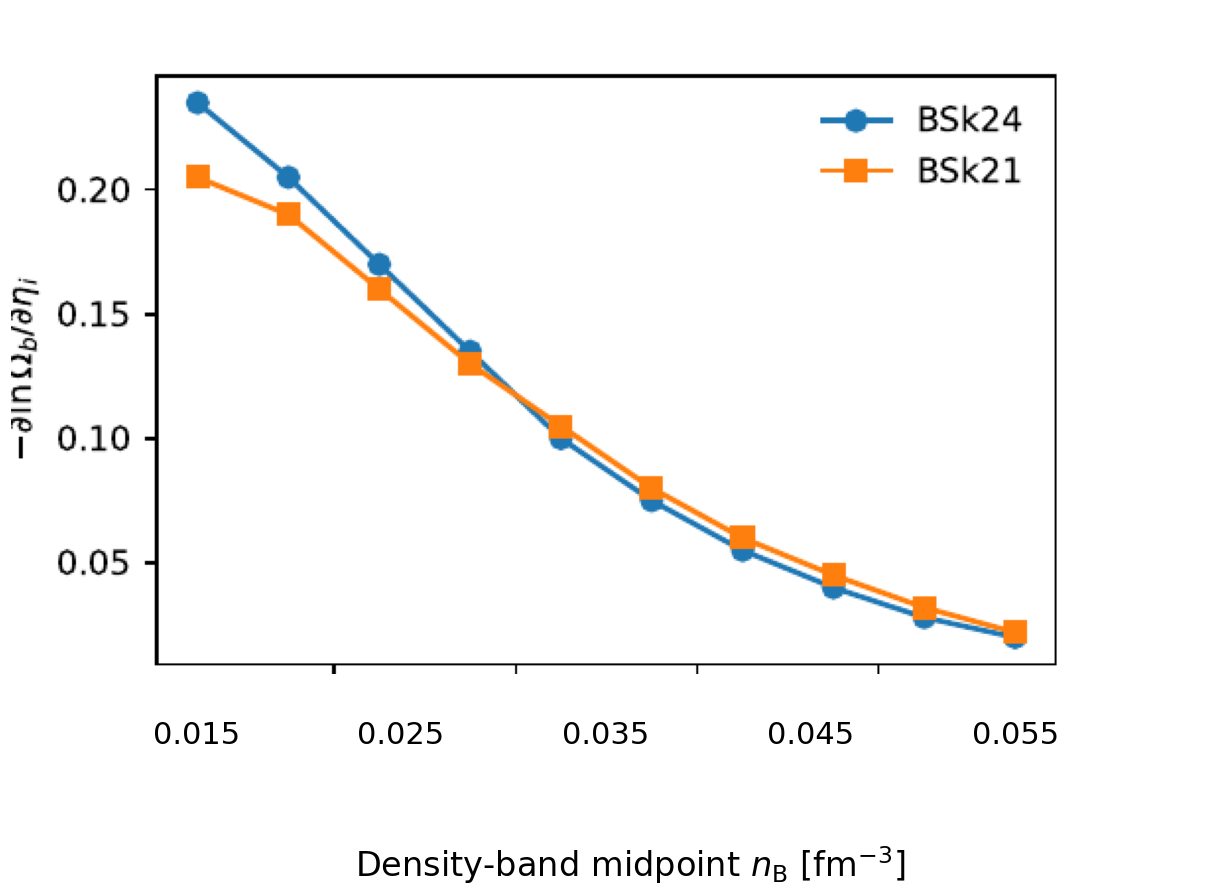}\caption{Density-resolved contribution of the physical frozen-composition correction. For the BSk models, most of the linear sensitivity comes from the outer part of the controlled inner-crust interval rather than from the deepest layer where the mechanical response alone is strongest.}\label{fig:physicalkernel}\end{figure}
The complementary experiment that isolates the purely mechanical susceptibility is shown in Appendix~\ref{app:kerneldiagnostics}. It rises inward and therefore has a different radial shape from Fig.~\ref{fig:physicalkernel}; their mismatch is precisely why the overlap, rather than either profile separately, controls the total failure shift.
\FloatBarrier

\section{Physical interpretation}\label{sec:interpret}
The calculations establish a simple chain of causation. Chemical freezing changes the local compressional stiffness; the elastic equations redistribute displacement between radial and tangential components; this changes the deviatoric strain field; and failure occurs at a different spin and, in general, a different depth. The roughly $20$ per cent change in $\Omega_b$ is therefore not proportional to a single average value of $K_f/K_{\rm eq}$. It emerges from a global response to a radially structured constitutive perturbation.

This clarifies the relation to \citet{Giliberti2020}. That work demonstrated, using a polytropic background and a phenomenological perturbative adiabatic index, that stratification can strongly alter rotation-induced strain. Here the same physical idea is tied directly to realistic inner-crust composition and turned into a prediction for the failure threshold. The EoS sets the radial source $K_f-K_{\rm eq}$, while the elastic problem determines the susceptibility with which each layer contributes to failure.

The result connects naturally to other crustal problems. Starquake models depend on where rotational stress accumulates \citep{BaymPines1971,FrancoLinkEpstein2000}; maximum-mountain calculations depend on how close different crustal layers are to yield \citep{Ushomirsky2000,HaskellJonesAndersson2006,JohnsonMcDanielOwen2013,GittinsAnderssonJones2021}; and magnetically or tidally loaded crusts likewise sample a non-uniform elastic susceptibility \citep{Lander2015,Tsang2012,GittinsAnderssonJones2021}. The overlap picture therefore suggests a broader lesson: when composition is frozen, replacing a stratified response by a single effective adiabatic index can miss not only the magnitude but also the location of the mechanically dominant region.

The mass sequence in Fig.~\ref{fig:massscan} is part of the physical result: it shows that the relative frozen penalty remains almost unchanged while the absolute failure scale varies. The remaining supporting tests are collected in Appendix~\ref{app:robust}. They show that the sign and order of the frozen correction survive changes in shear prescription, inner boundary and failure diagnostic, but these checks are not the central result.

\section{When is the frozen limit relevant?}\label{sec:timescales}
Equilibrium and frozen composition are limiting constitutive responses rather than different stellar backgrounds. The control parameter is schematically $\omega_{\rm load}\tau_\beta$: reactions maintain local equilibrium when $\omega_{\rm load}\tau_\beta\ll1$, while composition is effectively frozen when $\omega_{\rm load}\tau_\beta\gg1$. This is the same separation of timescales that underlies compositionally stratified oscillation calculations \citep{ReiseneggerGoldreich1992}.

The important point here is not to assign every astrophysical event to one limit, but to identify what the two limits mean mechanically. A rapid event probes $K_f$; sufficiently slow secular evolution probes $K_{\rm eq}$; an intermediate process requires a finite-reaction-time constitutive model. The illustrative reaction estimates of \citet{YakovlevGusakovHaensel2018} show that the regime can vary strongly with temperature, density and pairing. A unique crossover temperature cannot be quoted without evaluating the relevant reaction rates throughout the crust.

We therefore do not introduce an ad hoc interpolation between $K_{\rm eq}$ and $K_f$ here. Such an interpolation would require a frequency- and density-dependent complex response and belongs to a finite-reaction-time extension. The current calculation establishes the two mechanically well-defined endpoints and quantifies the full frozen correction with realistic microphysics.

The model remains Newtonian and uses one-way self-gravity. It treats the inner crust as an effective single elastic medium and omits explicit entrainment dynamics, plastic flow, fracture propagation and a relativistic rotating background. Relativity will modify crustal stratification and mechanical susceptibility, especially near the crust base \citep{JohnsonMcDanielOwen2013}; it does not determine the sign of the differential frozen/equilibrium correction without solving the relativistic boundary-value problem. These limitations matter more for an absolute breaking frequency than for the matched constitutive comparison presented here.

\section{Conclusions}\label{sec:conclusions}
The response of a neutron-star crust to rotation is controlled by more than the shear modulus and the equilibrium EoS. If composition cannot readjust during loading, the relevant compressional stiffness is $K_f$ rather than $K_{\rm eq}$. Using realistic BSk24, BSk21 and SLy4 microphysics, we find that this constitutive change lowers the rotational failure spin by $16.7$--$21.7$ per cent in the matched inner-crust domain. For BSk24 the reduction remains $21.05$--$21.97$ per cent from $1.2$ to $2.0\,M_\odot$, even though the absolute Newtonian failure frequency decreases across the sequence.

The main physical result is spatial. In BSk24 the first-failure layer shifts from $n_{\rm B}\simeq0.0284$ to $0.0361~{\rm fm^{-3}}$, while the direction and size of the migration vary with the EoS. Density-resolved experiments show why: the global failure shift is controlled by the overlap between the radial profile of frozen-composition stiffening and the mechanical susceptibility of the crust. For BSk21/24, about $82$ per cent of the linear sensitivity arises from $0.01<n_{\rm B}<0.03~{\rm fm^{-3}}$. A peak value or average adiabatic index therefore cannot by itself predict the response.

This work therefore provides a microphysical continuation of the earlier phenomenological stratification treatment: the result that a non-equilibrium adiabatic response can strongly reshape rotational strain is now tied to realistic composition profiles and to a failure threshold. The broader implication is that rapidly loaded neutron-star crusts should be treated as compositionally stratified elastic media. A natural next step is a finite-reaction-time calculation connecting the equilibrium and frozen endpoints as a function of density, temperature and loading timescale.

\section*{Data availability}
The numerical scripts and processed data underlying this article, including the microphysical reconstruction and numerical validation products, are available from the corresponding author on reasonable request.

\appendix
\section{Radial hydro-elastic system}\label{app:odes}
For reproducibility we give the strong-form system implemented in the numerical solver. The code uses $G=M=R=1$; radii are measured in units of $R$, densities in $M/R^3$, pressures and tractions in $GM^2/R^4$, potentials in $GM/R$, and angular frequencies in $(GM/R^3)^{1/2}$. Primes denote derivatives with respect to the dimensionless radius. No empirical normalisation is applied to any displacement, traction, or strain variable.

\subsection{Fluid rotational potential}
The rotational forcing used by all elastic closures is obtained first from the equilibrium fluid problem. Defining
\begin{equation}
W_\ell=\delta\Phi_\ell+\chi_\ell,
\end{equation}
and $L\equiv\ell(\ell+1)$, the radial equation solved by the code is
\begin{equation}
W_\ell''+\frac{2}{r}W_\ell'-\frac{L}{r^2}W_\ell+\frac{4\pi\rho}{c_e^2}W_\ell=-2\omega^2\delta_{\ell0},
\label{eq:appW}
\end{equation}
where $c_e^2=\Gamma_{\rm eq}P/\rho$ and $\omega^2\equiv\Delta\Omega^2R^3/(GM)$. The centrifugal-potential components are
\begin{equation}
\chi_0=-\frac{\omega^2r^2}{3},\qquad \chi_2=+\frac{\omega^2r^2}{3}.
\end{equation}
Regularity at the centre gives $W_0'(0)=0$ and $W_2'-2W_2/r\rightarrow0$. Matching the gravitational-potential perturbation to the exterior vacuum solution gives, at $r=1$,
\begin{equation}
W_\ell'+(\ell+1)W_\ell=(\ell+3)\chi_\ell.
\label{eq:appWbc}
\end{equation}
The same $W_0$ and $W_2$ are supplied to the equilibrium and frozen elastic calculations, so changing the compressional closure does not change the imposed rotational forcing.

\subsection{Definitions of the elastic variables}
For either multipole,
\begin{equation}
\xi_r=U_\ell(r)P_\ell(\cos\theta),\qquad
\xi_\theta=V_\ell(r)\,\partial_\theta P_\ell(\cos\theta),
\end{equation}
with $V_0=0$. Define
\begin{equation}
A_\ell\equiv\frac{2U_\ell-LV_\ell}{r},\qquad
D_\ell\equiv U_\ell'+A_\ell=\nabla\!\cdot\!\boldsymbol{\xi}_\ell,
\label{eq:appAD}
\end{equation}
and
\begin{equation}
\delta\rho_\ell=-\rho D_\ell-\rho'U_\ell.
\label{eq:appdrho}
\end{equation}
The bulk modulus is $K=\Gamma P$, with $\Gamma=\Gamma_{\rm eq}$ or $\Gamma_f$. We further define
\begin{equation}
\lambda=K-\frac{2}{3}\mu,\qquad \beta=K+\frac{4}{3}\mu.
\label{eq:applame}
\end{equation}
The radial and tangential traction amplitudes used as integration variables are
\begin{align}
T_\ell&=\beta U_\ell'+\lambda A_\ell-\rho gU_\ell,\label{eq:appTdef}\\
S_\ell&=\mu\left[V_\ell'+\frac{U_\ell-V_\ell}{r}\right],\label{eq:appSdef}
\end{align}
where $S_0=0$. For the quadrupole system we also introduce
\begin{equation}
Q_\ell\equiv\lambda D_\ell-\rho gU_\ell.
\label{eq:appQ}
\end{equation}
Equations~(\ref{eq:appAD})--(\ref{eq:appQ}) make the dependence on the constitutive input $K(r)$ explicit; no unspecified functions $F_R$ or $F_S$ remain.

\subsection{Monopole system}
Here and below $W_0$ retains the full fluid forcing defined in equation~(A1), $W_0=\delta\Phi_0+\chi_0$; it is not the centrifugal potential alone. For $\ell=0$, $L=0$ and $A_0=2U_0/r$. The solver integrates
\begin{align}
U_0'&=\frac{T_0-\lambda A_0+\rho gU_0}{\beta},\label{eq:appU0}\\
T_0'&=-\frac{2\mu}{r}\left(2U_0'-A_0\right)+g\,\delta\rho_0+\rho W_0'.\label{eq:appT0}
\end{align}
At the inner and outer fluid--solid interfaces, $r_b$ and $r_t$,
\begin{equation}
T_0(r_b)=\rho(r_b)W_0(r_b),\qquad T_0(r_t)=\rho(r_t)W_0(r_t).
\label{eq:appbc0}
\end{equation}

\subsection{Quadrupole system}
Likewise, $W_2=\delta\Phi_2+\chi_2$ includes the fluid gravitational response as well as the quadrupolar centrifugal forcing. For $\ell=2$, $L=6$. The four first-order equations solved for $(U_2,V_2,T_2,S_2)$ are
\begin{align}
U_2'&=\frac{T_2-\lambda A_2+\rho gU_2}{\beta},\label{eq:appU2}\\
V_2'&=\frac{S_2}{\mu}-\frac{U_2-V_2}{r},\label{eq:appV2}\\
T_2'&=\frac{LS_2}{r}-\frac{2\mu}{r}\left(2U_2'-A_2\right)+g\,\delta\rho_2+\rho W_2',\label{eq:appT2}\\
S_2'&=\frac{\rho W_2}{r}-\frac{1}{r}\left[3S_2+Q_2+\frac{2\mu U_2}{r}+\frac{2\mu(1-L)V_2}{r}\right].\label{eq:appS2}
\end{align}
The boundary conditions are
\begin{align}
T_2(r_b)&=\rho(r_b)W_2(r_b), & S_2(r_b)&=0,\\
T_2(r_t)&=\rho(r_t)W_2(r_t), & S_2(r_t)&=0.
\label{eq:appbc2}
\end{align}
These are precisely the boundary conditions used in every matched equilibrium/frozen comparison reported in the main text.

\subsection{Strain reconstruction}
The strain invariant is reconstructed directly from the displacement solution rather than from the traction variables. For one multipole the non-zero coordinate components are
\begin{align}
\varepsilon_{rr}^{(\ell)}&=U_\ell'P_\ell,\\
\varepsilon_{r\theta}^{(\ell)}&=\frac12\left(V_\ell'-\frac{V_\ell}{r}+\frac{U_\ell}{r}\right)\partial_\theta P_\ell,\\
\varepsilon_{\theta\theta}^{(\ell)}&=\frac{U_\ell}{r}P_\ell+\frac{V_\ell}{r}\partial_\theta^2P_\ell,\\
\varepsilon_{\phi\phi}^{(\ell)}&=\frac{U_\ell}{r}P_\ell+\frac{V_\ell}{r}\cot\theta\,\partial_\theta P_\ell.
\label{eq:appstrain}
\end{align}
The physical field is the sum of the $\ell=0$ and $\ell=2$ contributions. Its trace is removed before evaluating the von-Mises invariant. For $\ell=2$, the implementation uses $P_2=(3\cos^2\theta-1)/2$, $\partial_\theta P_2=-3\sin\theta\cos\theta$, and $\cot\theta\,\partial_\theta P_2=-3\cos^2\theta$, including their regular polar limits.

\section{Numerical implementation and microphysical audit}\label{app:numerics}
Equilibrium and frozen models are generated by the same boundary-value solver with identical background, forcing, fluid gravitational response, boundaries and shear modulus. Unless stated otherwise we use Newtonian $1.4\,M_\odot$ backgrounds and an elastic shell from $n_{\rm B}=0.06~{\rm fm^{-3}}$ to $\rho=10^{12}~{\rm g\,cm^{-3}}$; the failure diagnostic is restricted to the controlled $0.01$--$0.06~{\rm fm^{-3}}$ interval. Production solutions are evaluated at $\Omega_*=0.5\Omega_K$ and rescaled to failure.

At each density the frozen closure is reconstructed from the local Wigner--Seitz composition. Schematically,
\begin{equation}
K_{f,{\rm raw}}=\left.\frac{dP(\lambda n_a)}{d\ln\lambda}\right|_{\lambda=1}.
\end{equation}
For all three EoSs we use the common pressure-renormalised form $K_f=\Gamma_{f,{\rm edge}}P_{\rm eq}$ together with the conservative regularisation $K_f\ge K_{\rm eq}$. The archived production products do not retain the pre-floor point-by-point arrays needed to quote a certified activation fraction. We therefore do not infer an activation geometry from post-floor curves. For BSk24, an equivalent pressure-renormalised implementation changes the failure reduction by less than $4\times10^{-5}$ percentage points, showing that the principal result is insensitive to this implementation detail at the quoted precision.

A separate BSk21 microphysical audit gives a median inner-crust value $K_f/K_{\rm eq}\simeq1.095$ and a deep-layer value approaching $1.39$. These scalar diagnostics refer to that independent BSk21 audit and should not be read off Fig.~\ref{fig:kf}, which intentionally displays only the pointwise archived BSk24 profile. The failure calculation uses the EoS-specific radial closure rather than either scalar summary.

The production baseline uses the Voigt Coulomb-lattice modulus. A physically motivated polycrystalline sensitivity adopts the self-consistent result of \citet{KobyakovPethick2015},
\begin{equation}
\mu_{\rm poly}/\mu_V=0.3778/0.4852=0.77865.
\end{equation}
An independent regression changes its own BSk24 frozen reduction from $21.55$ to $22.46$ per cent. Applying only the computed poly/Voigt response factors to the certified baseline gives $22.62$ per cent; this is a response-transfer diagnostic, not a separate production rerun. Both are summarised as a shear-law sensitivity of about $22.5$ per cent.

\section{Density-resolved and summary diagnostics}\label{app:kerneldiagnostics}
The physical-kernel experiment in Section~\ref{sec:results} changes only the actual frozen-minus-equilibrium constitutive correction in one density band at a time. A complementary diagnostic instead applies the same small fractional increase in $\Gamma$ to each density band. This removes the EoS-dependent amplitude of the frozen correction and exposes the intrinsic mechanical susceptibility of the boundary-value problem.
\begin{figure}
\centering
\includegraphics[width=\columnwidth]{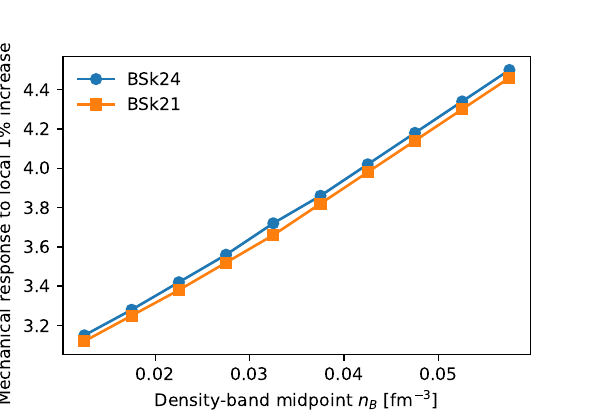}
\caption{Purely mechanical sensitivity to the same local fractional increase in $\Gamma$. The susceptibility grows inward, unlike the physical frozen-composition contribution shown in Fig.~\ref{fig:physicalkernel}; the global failure shift depends on the radial overlap of the two.}
\end{figure}

\begin{figure}
\centering
\includegraphics[width=\columnwidth]{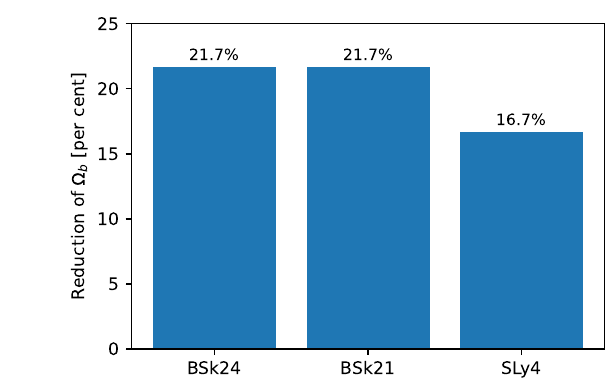}
\caption{Summary reduction of the rotational failure frequency in the matched $1.4\,M_\odot$ bulk comparison. This plot is retained as an auxiliary numerical summary; the main text emphasises the strain redistribution, mass dependence and radial-overlap physics.}
\label{fig:eosreduction}
\end{figure}

For completeness, Fig.~\ref{fig:eosreduction} gives the compact EoS-by-EoS summary of the matched-bulk reduction. The corresponding absolute frequencies and ratios are reported in Table~\ref{tab:eos} in the main text.

\section{Robustness and auxiliary scans}\label{app:robust}
The following calculations test whether the main frozen-composition effect is tied to one mechanical prescription. They are intentionally separated from the main physical narrative.
\begin{table}
\centering
\caption{Separate BSk24 sensitivity tests. They are not a combined uncertainty budget.}
\begin{tabular}{lc}
\toprule variation&reduction\\
\midrule
certified Coulomb--Voigt, von Mises&21.71\%\\
$\mu=10^{16}\rho$, von Mises&21.66\%\\
self-consistent polycrystal (audit)&$\simeq22.5$\%\\
physical crust--core boundary, von Mises&15.30\%\\
matched bulk, Tresca&26.48\%\\
\bottomrule
\end{tabular}
\end{table}
The separately certified BSk24 variations span $15.3$--$26.5$ per cent. Because the physical crust--core boundary and Tresca variations have not been applied simultaneously, this is a robustness envelope rather than a propagated uncertainty. For BSk21 the physical crust--core boundary changes the reduction from $21.66$ to $17.03$ per cent.

\begin{table}
\centering
\caption{Numerical values underlying the BSk24 mass sequence shown in Fig.~\ref{fig:massscan}.}
\label{tab:mass}
\begin{tabular}{lccc}
\toprule $M/M_\odot$&$\Omega_b^{\rm eq}/\Omega_K$&$\Omega_b^f/\Omega_K$&reduction\\
\midrule
1.2&0.5044&0.3982&21.05\%\\
1.4&0.4997&0.3912&21.71\%\\
1.6&0.4919&0.3864&21.45\%\\
1.8&0.4899&0.3823&21.97\%\\
2.0&0.4847&0.3786&21.89\%\\
\bottomrule
\end{tabular}
\end{table}
These values underlie the main-text mass-dependence result. The nearly constant differential reduction is the robust feature; the absolute Newtonian sequence is not used to infer a quantitative relativistic mass trend.

\section{Reproducibility gates}
The production workflow enforces four gates: (i) the same background, $W_\ell$, boundaries and $\mu$ are used in each equilibrium/frozen pair; (ii) $\eta=0$ and $\eta=1$ recover the equilibrium and full-frozen solvers in kernel experiments; (iii) grid refinement changes the principal maximum by less than $10^{-3}$ relative, with the BSk24 invariant audit below $10^{-5}$; and (iv) the microphysical frozen closure is used only in the directly controlled $0.01$--$0.06~{\rm fm^{-3}}$ interval and is smoothly returned to equilibrium outside it. The polycrystalline calculation is explicitly labelled an independent regression sensitivity and does not replace the certified Voigt production table.

\label{lastpage}\end{document}